# Strength and Stability Analysis of a Single Walled Black Phosphorus Tube under Axial Compression

Kun Cai [1, 2], Jing Wan [1], Ning Wei [1], Qinghua Qin [2*]

[1] *College of Water Resources and Architectural Engineering, Northwest A&F University, Yangling 712100, China*
[2] *Research School of Engineering, the Australian National University, Canberra, ACT, 2601, Australia*
[*] Corresponding author's email address: qinghua.qin@anu.edu.au (Qinghua Qin)

Abstract

Few-layered black phosphorus materials recently attract much attention due to its special electronic properties. As a Consequence, the nanotube from a single-layer black phosphorus (SLBP) has been theoretically built. The corresponding lectronic properties of such black phosphorus nanotube (BPNT) were also evaluated numerically. However, unlike the graphene formed with 2sp2 covalent carbon atoms, a SLBP is formed with $3sp^3$ bonded atoms. It means that the structure from SLBPwill possess lower Young's modulus and mechanical strengths than those of carbon nanotubes. In this study, molecular dynamics simulation is performed to investigate the strength and stability of BPNTs affected by the factors of diameter, length, loading speed and temperature. Results are fundamental for investigating the other physical properties of a BPNT acting as a component in a nanodevice. For example, the buckling of BPNT happens earlier than the fracture, before which the nanostructure has very small axial strain. For the same BPNT, higher load speed results in lower critical axial strain and the nanotube with lower axial strain can still be in stable state at a higher temperature.

## 1. Introduction

In recent years, as a new two-dimensional semiconductor material, few-layer black phosphorus behaves excellent electronic properties [1-4] as comparing with graphene [5, 6] boron nitride [7, 8] and $MoS_2$ [9, 10]. Considering such excellent material properties, much effort has been taken on the research of its potential applications. Unlike graphene which is formed with 2sp2 carbon atoms, a single-layer black phosphorus (SLBP or phosphorene) is formed by bonding phosphorus atoms together with 3sp3 hybridized bonds. And the interlayer interaction of many-layered black phosphorus is mainly due to the van der Waals attraction [11]. Hence, the in-plane strength and stiffness of a SLBP is not as high as those of the graphene [6]. Obviously, as a component containing SLBP in a nanodevice, the strength and stability of the structure is fundamental to ensure the normal operation of the device. The mechanical properties of a SLBP have been investigated much deeply during the past decade. For example, on testing the in-plane moduli of SLBP along different directions, Jiang and Park [12] claimed that the elastic modulus is ~106.4GPa along the pucker direction and ~41.3 GPa in its vertical direction. They also discovered [13] that the out-of-plane Poisson's ratio of SLBP is negative when a SLBP is under uniaxial tension along pucker direction. Whereas the modulus reported by Wei and Peng [14] is ~44 GPa only. In the following work, the ideal tensile strength and critical strain of SLBP were discussed. Wang *et al.* [15] obtained the same value of the modulus in their study on the dynamics behavior of a nano resonator made of black phosphorus. Besides elastic modulus, researchers also investigated such mechanical properties as the edge stress of a SLBP [16], the coupling of mechanical deformation and electrical properties [17], a SLBP under shear load [18], the fracture limits of a SLBP [19]. Hence, the mechanical property is important for designing nanodevice with black phosphorus sheets.

Similar to the formation of a carbon nanotube [20-23] by curling a piece of graphene ribbon along one direction, researchers also suggested that a nanotube could be made from a SLBP nanoribbon. Indeed, some properties of a black phosphorus nanotube (BPNT) have been estimated using numerical method [24]. It is, however, still necessary to demonstrate the relationship between the material properties of a nanostructure and its strength and stability. In our previous work [25], we have found that the BPNT formed by curling the SLBP along different directions (e.g., armchair/pucker direction and zigzag direction) has different stability behavior at finite temperature. In one word, even just thermal vibration of the atoms on a BPNT at high temperature may lead to collapsing of the nanostructure. In the present study, we focus on the strength and stability of a BPNT under axial compression. Some factors affecting the design of a future nanodevice with BPNT components, such as the slenderness, length, loading speed and temperature of environment are investigated.

2. Models and Method

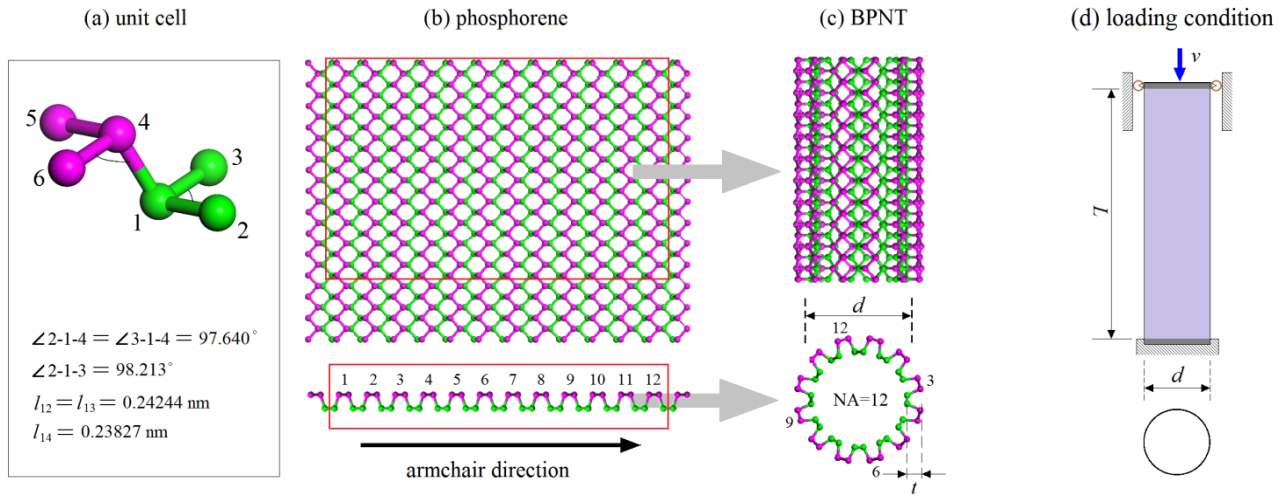

Figure 1. Geometrical model of a black phosphorous nanotube (BPNT). (a) The unit cell of a single layer black phosphorus (SLBP). (b) A nanoribbon of a single layer phosphorene. (c) The BPNT obtained by curling the ribbon within the red frame in (b) along the armchair direction. In the model, NA=12 means that there are 12 periodic unit cells along the circular direction (i.e., the curved armchair direction). "$d$" is the average diameter of the BPNT, i.e., half of the summation of the diameters of the two layers. "$t$" is the thickness of SLBP. (d) The loading condition for a BPNT under uniaxial compression. "$v$" is the velocity of the left end during loading. The grey areas of the tube are fixed on the degree of freedoms within the cross-section of the tube. In this study, the thickness of each grey area is 0.1657nm. "$L$" is the effective length of tube, which is 0.3314nm shorter than the total length of tube ($L^*$).

Figure 1 demonstrates a BPNT made from a SLBP by geometrical mapping method. Figure 1d presents loading conditions of a cylindrical tube with average diameter of "$d$" for the mechanical model used in the present simulation. The atoms within the grey area are fixed during simulation. Four factors are considered in the following analysis. **The first** one is the effect of the diameter of tube. The parameters of tubes involved are listed in Table 1. **The second** is the length effect. In discussing the length effect, the tubes with the same diameter (or NA=20, or $d$=2.785 nm) have different length. **The third** is the effect of loading speed (or value of $v$). In the discussion, the tube with NA=20 and L=20.166 nm is adopted. **The final** is the temperature effect. For the strength and stability of the BPNT with NA=20 and L=20.166 nm, the environmental temperature increases from 0.1 K to a the critical value at which the tube is either fractured or has global buckling. Details are given in the following sections.

The open code for molecular dynamics simulation [26] is used in the present study. Meanwhile, the Stillinger-Weber (S-W) potential developed by Jiang et al. [27] is used to model the interaction among atoms in a BPNT.

The major steps in molecular dynamics simulation here are as follows. **1)** Create a BPNT (e.g., with specified value of NA and $L$ or $L^*$). **2)** Reshape the tube by energy minimization using steepest descent algorithm. **3)** Take a thermal bath for the tube in the canonical NVT ensemble with T=0.1K for 50 ps. **4)** Initiate the velocities of atoms to be zero and fixed the atoms (in the grey areas in Figure 1d) at two ends of the tube. **5)** Keep the free atoms in the above NVT ensemble and, simultaneously, start displacement loading step. In each loading step, the upper end of the tube moves down of "$s$" within 1 time step (i.e., 1fs) but followed with "$t_R$" period of relaxation. **6)** Stop the loading step when buckling/fracture of the BPNT happens obviously.

Table 1 Parameters of BPNTs in discussion on the effect of diameter. $\alpha^*=L^*/d$, $L=L^*-0.3314$nm, $\alpha=L/d$.

|        | Model 1 | Model 2 | Model 3 | Model 4 | Model 5 |
|--------|---------|---------|---------|---------|---------|
| NA     | 14      | 16      | 18      | 20      | 22      |
| $d$/nm | 1.951   | 2.221   | 2.507   | 2.785   | 3.065   |
| $L^*$/nm | 15.576 | 17.896 | 20.050 | 22.204 | 24.524 |
| $\alpha^*$ | 7.9836 | 8.0576 | 7.9976 | 7.9727 | 8.0013 |
| $L$/nm | 15.245  | 17.565  | 19.719  | 21.873  | 24.193  |
| $\alpha$ | 7.8137 | 7.9084  | 7.8654  | 7.8537  | 7.8932  |

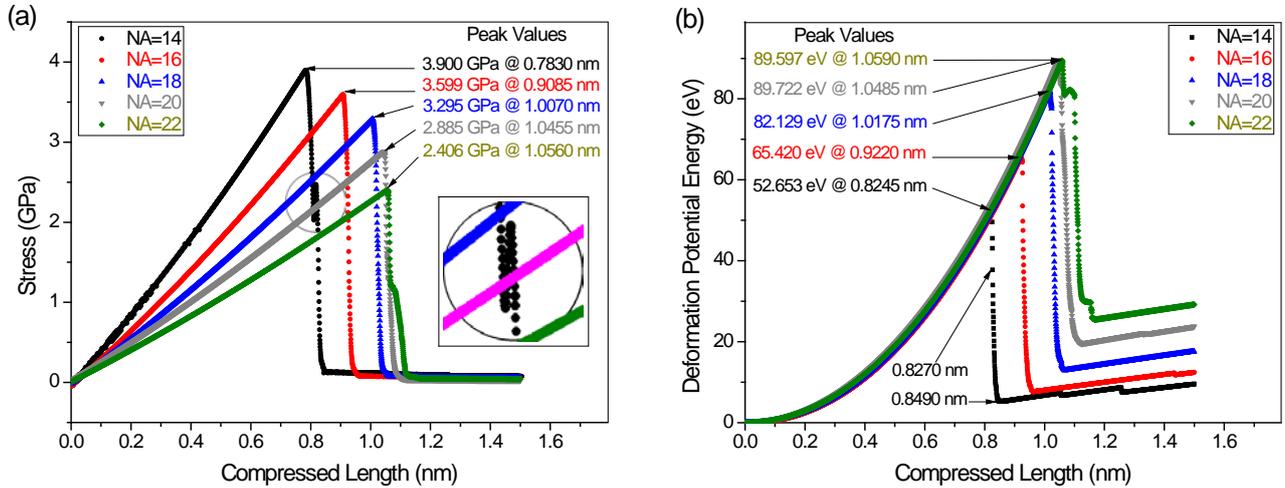

Figure 2. The stress and deformation potential energy of the BPNTs (with the same value of $\alpha$) under compression. (a) The stress v.s. compressed length curve. The stress is the average value of the axial force over the cross-sectional area of tube. (b) The deformation potential energy (DPE) v.s. compressed length curve. The initial potential energy of the five cases is -1385.55 (NA=14), -1831.95 (NA=16), -2320.17 (NA=18), -2864.74 (NA=20) and -3489.21 eV (NA=22), respectively.

3. Results and Discussion

*3.1 Effect of nanotube diameter on the strength and stability of BPNT*

In this section, five BPNTs are involved in simulation to demonstrate the diameter effect on the strength and stability of BPNT as they have similar ratio of length over diameter. The geometrical parameters of the tubes are listed in Table 1, e.g., NA=12, 14, 16, 18, 20, 22. The velocity of load is set by specifying *s*=0.0005nm and it is followed with 2 ps of relaxation (i.e., $t_R$=2 ps).

Due to the different initial lengths of BPNTs, only accumulation of the displacement load acts as the level axis in the following curve figures.

From Figure 2a, one can find that the maximal value of stress decreases with the increase of the diameter of tube (or NA). The reason can be explained from the Euler's buckling formulation for buckling of a linear elastic column [28, 29]:

$$\sigma_{cr} = \frac{4\pi^2 \cdot E \cdot I_x}{L^2 \cdot A} ,  \qquad (1)$$

where $\sigma_{cr}$ is the critical stress of a column under axial compression, "*E*" is the axial modulus of the material, "*A*" the cross sectional area of the bar, "$I_x$" is the moment of inertia with respect to x-axis. In the present model, "*A*" and "$I_x$" can be calculated according to the following formula.

$$A = \pi \cdot d \cdot t , \qquad (2)$$

$$I_x = \frac{1}{2} \int_0^{2\pi} d\theta \int_{(d-t)/2}^{(d+t)/2} r^3 dr = \frac{A}{8}(d^2 + t^2) , \qquad (3)$$

where "*t*" is the thickness of the tube, which is 0.2132 nm for SLBP. Hence, Eq. (1) can be modified as

$$\sigma_{cr} = \frac{\pi^2 \cdot E \cdot (d^2 + t^2)}{2L^2} = \frac{\pi^2}{2} E \cdot \left[ \left(\frac{1}{\alpha}\right)^2 + \left(\frac{t}{L}\right)^2 \right] . \qquad (4)$$

For the five BPNTs, the elastic moduli are identical. If, for example, the value of *α* is identical, too, the tube with lower length will have higher value of critical stress of buckling. It matches very well the results given in Figure 2a.

Making use of Eq. (4), the elastic modulus of a BPNT with axis normal to armchair direction can be calculated as 48.255 GPa (NA=14). The present modulus is very close to the results published in [12, 14, 15].

Figure 2b shows the potential energy due to deformation. The time at the peak value of deformation potential energy (DPE) is longer than that at the peak value of the stress. With the increase of NA, the difference between them tends to be zero. For example, when NA=14, the difference is 0.8245-0.7830 (=0.0415nm), which means 83 loads steps of difference. When NA=22, the difference is 1.0590-1.0560 (=0.0030nm), which is only 6 load steps of difference. In each curve of DPE, the value varies in three stages. Firstly, the value increases monotonously to the peak value. The structure deforms linearly in this period. Secondly, the value decreases (but may not monotonously) to the lowest value. This is because the tube is in the buckling state. Finally, as the structure is still in compression, some of the P-P bonds are broken (see Figure 3) and the deformation (curvature of tube wall) is increasing which leads to the increase of the DPE.

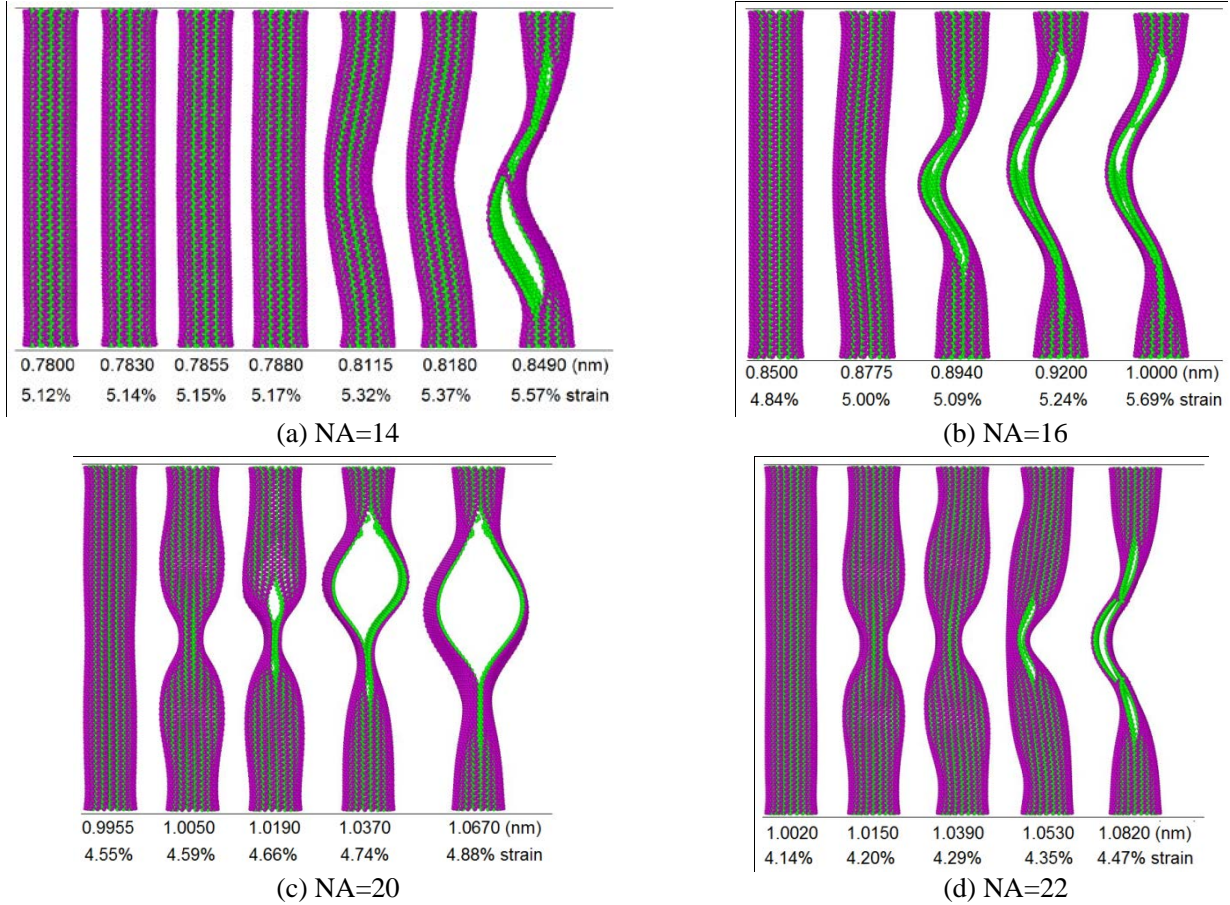

Figure 3. The configurations of the BPNTs during compression. For example, (a) for the tube with NA=14, the compressed length ($\Delta L$) is in [0.7800, 0.8490] nm. Strain is calculated using $\Delta L/L$. **The lower end of tube is fixed and the upper end moves down.**

There is a vibration for the transversal deformation of the BPNT with NA=14 during [0.7800, 0.8490] nm of compressed length (Movie 1). The vibration is also reflected in the stress history, as shown by the black data points circled in the Figure 2a. At 0.7830 nm of compressed length, the axial stress of the tube reaches its maximal value of 3.900 GPa. The stress drops to 2.0304 GPa at 0.8115 nm of compressed length. Then the stress increases to a local maximal value of 2.475 GPa at 0.8180 nm. Soon after that at 0.8490 ps the fracture of tube happens. As the maximal stress (or critical stress) approaches, the tube can be considered as in buckling state [28, 30, 31]. But this is only suitable for short tubes. When the length of BPNT is higher, the tube will be broken seriously at the critical stress state. For example, for the BPNT with NA=20, the stress reaches its peak value when the compressed length equals to 1.0450nm. From the third and the fourth configurations of the BPNT with NA=20 in Figure 3c, one can find that the BPNT has been broken seriously (Movie 2). Hence, the maximal stress of the tube cannot be considered as the critical stress of the first-ordered buckling.

In the post-buckling state [31-33], the fracture of the tube is due to bonds breakage as the curvature of the outer wall of BPNT is too large [25]. In other words, the stability of BPNT loses before the critical state of strength of the tube. The configurations in Figure 3 demonstrate that the critical strain with respect to the strength of a tube decreases with the increasing of the diameter. Besides, the deformation of the tubes before fracture is very small (e.g., 0.8490 nm is only 5.57% of the effective length of tube with NA=14), which indicates that the tubes are in the linear elastic deformation state. Hence, Eq. (1) is suitable for the current models.

## 3.2 Length effect on the strength and stability of BPNT

To demonstrate the influence of the tube length on the strength and stability, in this section, the BPNTs with the NA=20 ($d$=2.785nm) are adopted in the simulation. Four values of tube length are considered: $\alpha = L/d$=6, 8, 10 and 12. When $\alpha$ =8, the model is identical to the Model 4 (with NA=20) in Table 1. The major steps in simulation are similar with those mentioned above. The only difference is that the velocity of loading is setting with s=0.001nm followed with 2 ps of relaxation (i.e., $t_R$=2 ps).

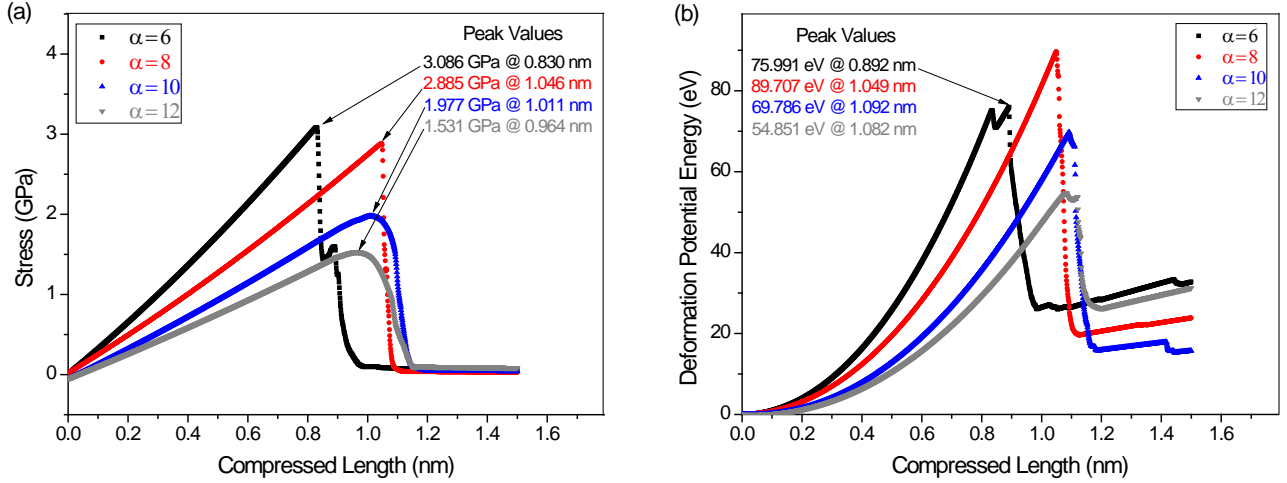

Figure 4. The stress and deformation potential energy of the BPNTs (with the same NA=20) under compression. (a) The stress-compressed length curve. (b) The deformation potential energy-compressed length curve. The initial potential energy of the four cases is -2161.28 ($\alpha$=6), -2864.74 ($\alpha$=8), -3589.48 ($\alpha$=10) and -4314.25 eV ($\alpha$=12), respectively.

It can be seen that from Figure 4a, that the critical stress (the peak value of stress) is higher for a shorter tube under compression. It agrees well with that from Eq. (4). However, the time when the stress approaches the peak value does not obey the rule well. Before buckling, the tube is in a linear elastic state. Hence, the critical strain can be expressed as

$$\varepsilon_{cr} = \frac{\sigma_{cr}}{E} = \frac{\pi^2 \cdot (d^2+t^2)}{2L^2} \quad . \tag{5}$$

From the definition of strain of a column, the critical strain can be expressed as

$$\varepsilon_{cr} = \frac{\Delta L_{cr}}{L} \quad , \tag{6}$$

where $\Delta L_{cr}$ is the critical value of compressed length of BPNT. Hence, we have

$$\Delta L_{cr} = \frac{\pi^2 \cdot (d^2+t^2)}{2L} \quad . \tag{7}$$

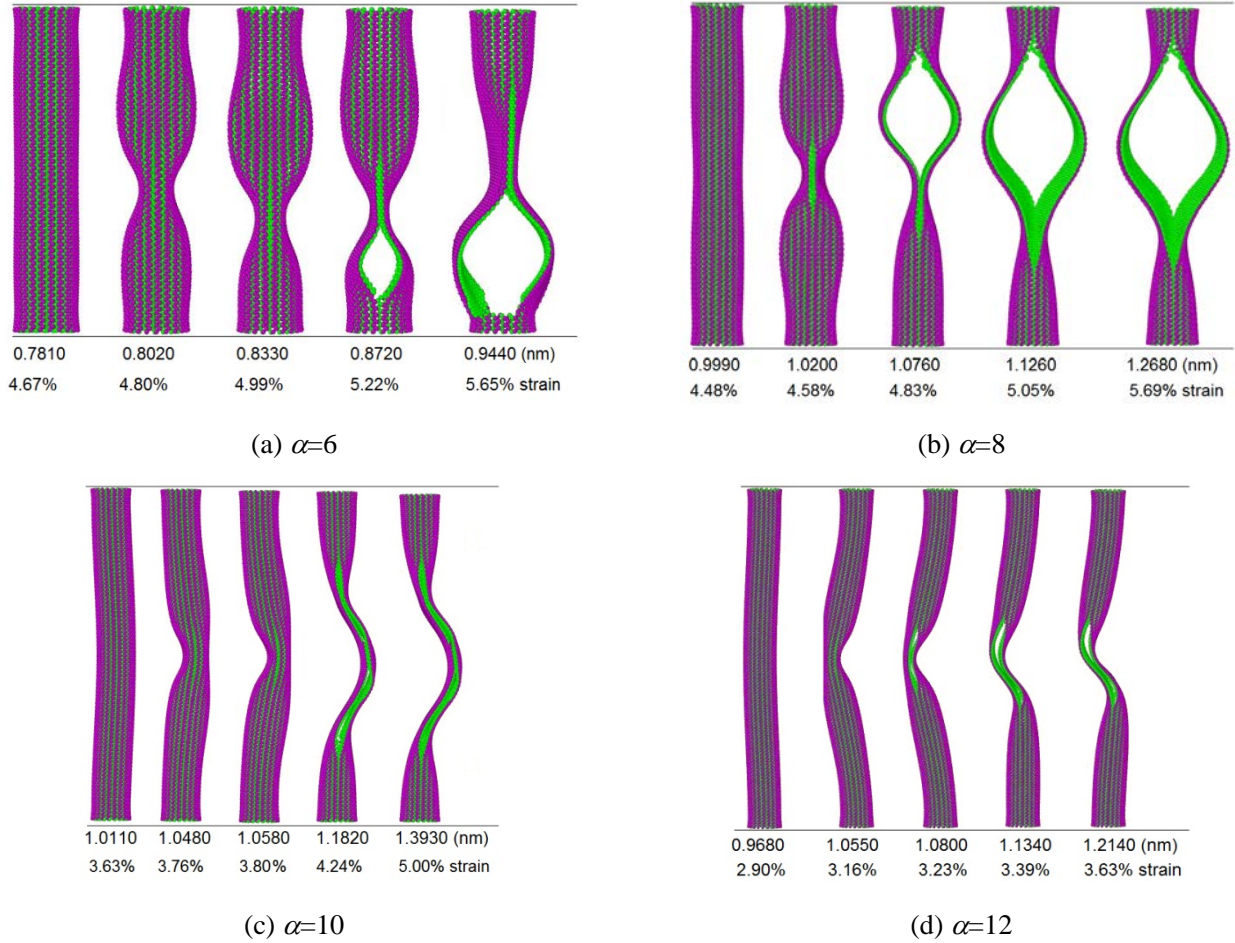

Figure 5. Representative configurations of BPNTs with different lengths during compression ($\alpha=L/d$).

The results shown in Figure 4 are obtained by investigating the BPNT with the same cross section (i.e., the same values of d and t) but different lengths under the same loading condition. Hence, Eq. (7) indicates that the critical value of the compressed length of BPNT is reversely proportional to its length. But this is not satisfied for $\alpha$ =6 (the shortest tube). When $\alpha$ =6, the critical compressed length is the smallest rather than the greatest among the four cases, because the critical stress appears earlier than the critical strain for the shortest tube.

Figure 4b shows the DPE of the tubes with different length, in which the peak value of the DPE of the tube with $\alpha$ =6 is lower than that of the tube with $\alpha$ =8. The reason is that the tube with $\alpha$ =6 is shorter and the compressed length is also lower than that of the tube with $\alpha$ =8. The DPE of the tube with $\alpha$ =8 is also higher than those of the tubes with $\alpha$ =10 and 12. The reason is that the tube of $\alpha$ =8 has higher value of critical stress and larger compressed length.

Figure 5 gives five representative configurations of the tube under compression. Comparing with Figure 4a, the peak value of stress appears before fracture starts. Commonly, the tubes have been in post-buckling state when their stress approaches maximum. One can also find that the critical strain with respect to the strength decreases with an increase in the length. For example, the critical strain is between (4.99%, 5.22%) when $L=6d$ ($\alpha$ =6, Figure 5a) (Movie 3). When $L=8d$ ($\alpha$ =8, Figure 5b), the critical strain is between (4.58%, 4.83%). If the length of tube is 12 times of the diameter, the critical strain is lower than 3.23% (the third configuration in Figure 5d).

*3.3 Effect of load speed on the strength and stability of BPNTs*

In this section, the effect of loading speed on the strength and stability of BPNT is discussed. The BPNT with NA=20 ($d$=2.785nm) and $L$=20.166 nm is adopted in the simulation. All major steps in our simulation are identical to those described in Section 3.1 except for loading step. Four types of loading speed (or relaxation schemes) are considered, i.e., $s$= 0.0005nm followed with $t_R$ = 0.2, 0.4, 1.0 and 2.0 ps of relaxation in each load step, or respectively, the loading velocities are ~ 2.5, 1.25, 0.5 and 0.25m/s.

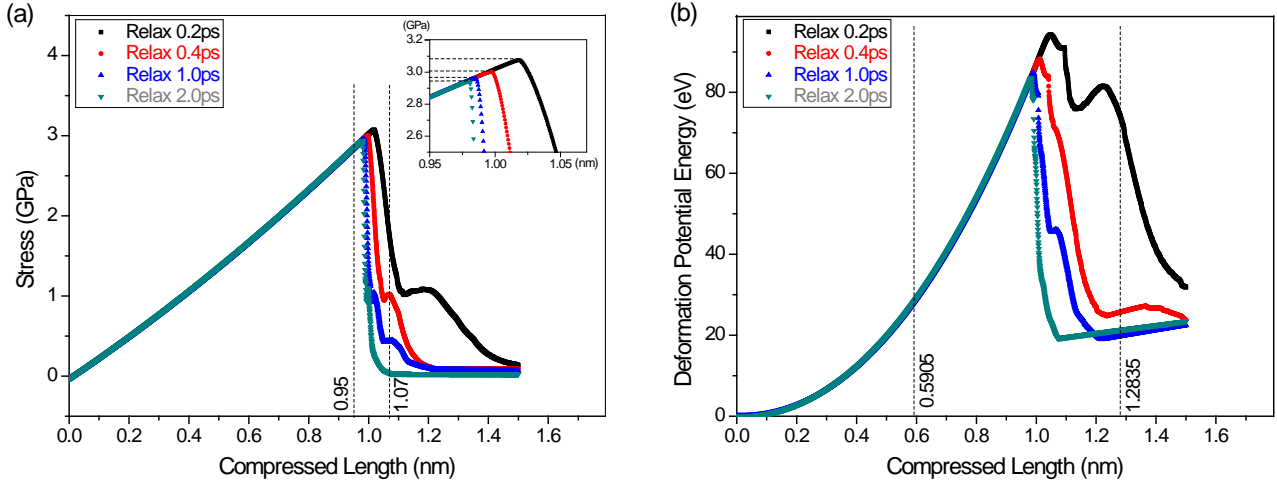

Figure 6. The stress and deformation potential energy of the same BPNT (NA=20, $L$=20.166 nm) under compression with different load speed, i.e., in each load step the relaxation time is 0.2, 0.4, 1.0 and 2.0 ps. (a) The stress-compressed length curve. The peak values are 3.075 GPa at 1.0175nm (relax 0.2ps), 3.002 GPa at 0.9970nm (relax 0.4ps), 2.960 GPa at 0.9850 nm (relax 1.0ps) and 2.9433 GPa at 0.9800 nm (relax 2.0ps), respectively. (b) The deformation potential energy-compressed length curve. The initial potential energy of the four cases is -587.59 eV.

Figure 6a indicates that the peak value of the critical stress of the BPNT is higher at higher loading speed. The stress gains its highest value when there is only 0.2ps of relaxation in each load step. But the differences among the peak values are not obvious. Figure 6b shows that the tube under the first relaxation scheme has the highest potential energy, too. The reason is that the tube has higher value of compressed length before buckling or fracture.

Figure 7 gives five snapshots of the tube under the load with different lengths of relaxation in each step for each case. Clearly, the tube has been in buckling state when the compressed length is 0.5905 nm for $t_R$=0.2ps. When the compressed length reaches 0.8240 nm, the tube is fractured. After that, the tube collapses, In this case, the configuration is with respect to 1.0160 nm of compressed length or 5.04% of engineering strain (Figure 7a) (Movie 4). **For the four cases, the strain of the BPNT, which is still at post-buckling state but not damaged, is higher at lower loading speed**, e.g., 2.93% ($t_R$ = 0.2ps) < 3.95% ($t_R$ = 0.4ps) < 4.61% ($t_R$ = 1.0ps) < 4.63% ($t_R$ = 2.0ps).

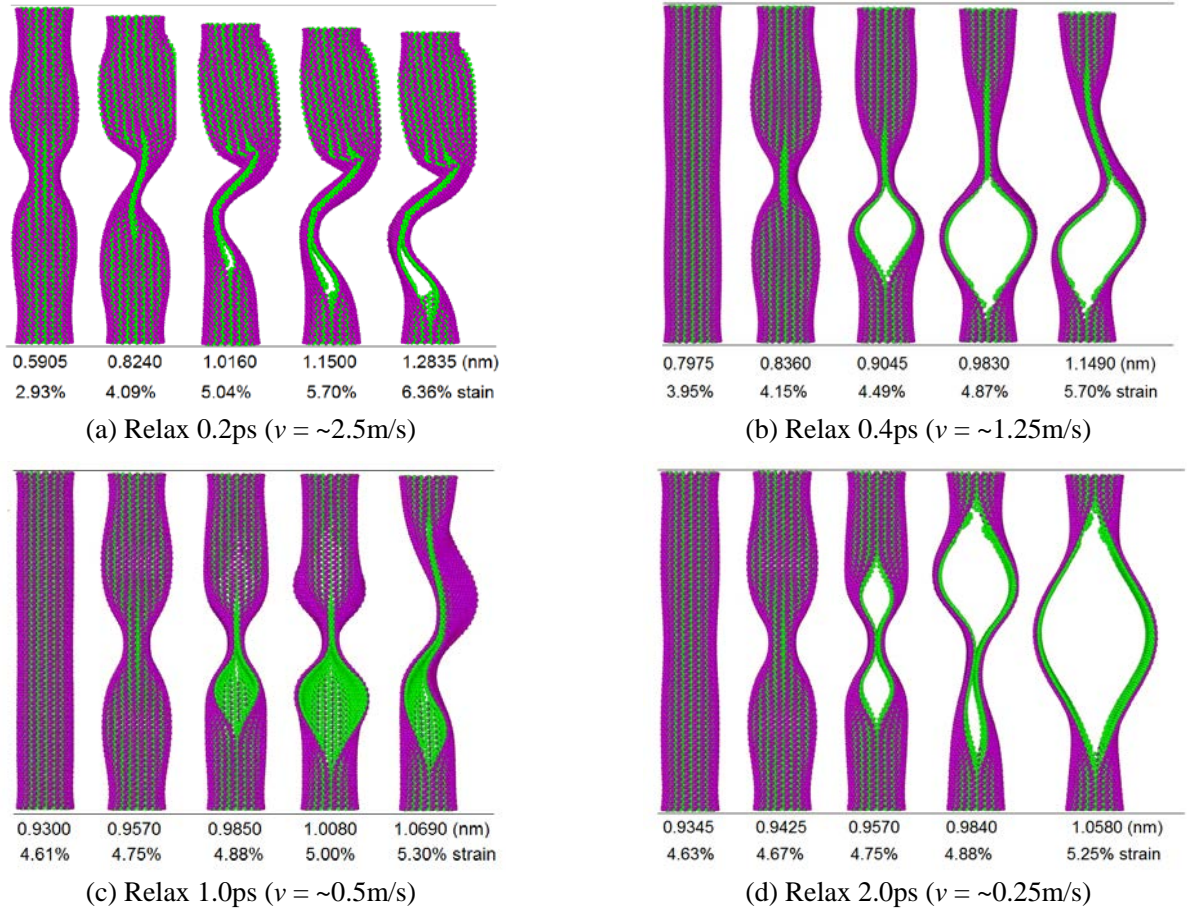

Figure 7. Representative snapshots of the BPNT (NA=20, $L$=20.166 nm) during compression with different loading speeds.

### 3.4 Effect of temperature on the strength and stability of BPNTs

In the discussion above, the temperature is set to be 0.1K. It means that the results obtained have nothing to do with the thermal vibration of atoms on tubes. Actually, a device is commonly running at different temperature. Hence, temperature is also an important factor in affecting the strength and stability of BPNTs. The major steps in our simulation are much similar to those described above. The only difference is that the environmental temperature will increase from 0.1K with the speed of 1K/ns.

The BPNT used in this section is the same as that in Section 3.3, in which NA=20, $L$=20.166 nm. From the discussion above, we know that the tube buckles when the compressed length approaches 0.9800nm (relaxes for 2ps in each load step). As shown in our previous study [25], stronger thermal vibration of the atoms on the BPNT leads to failure of structure easier. Hence, we believe that the compressed length will be shorter at higher temperature for the same BPNT. In the present simulation, we choose the maximal compressed length of the BPNT being 0.1, 0.3, 0.5, 0.7 and 0.9 times of the 0.9800 nm. In each loading step, $s$=0.0005nm followed with 2.0ps of relaxation. When the compressed length reaches the specified value, the environmental temperature starts to increase until the tube is at buckling state or fractured.

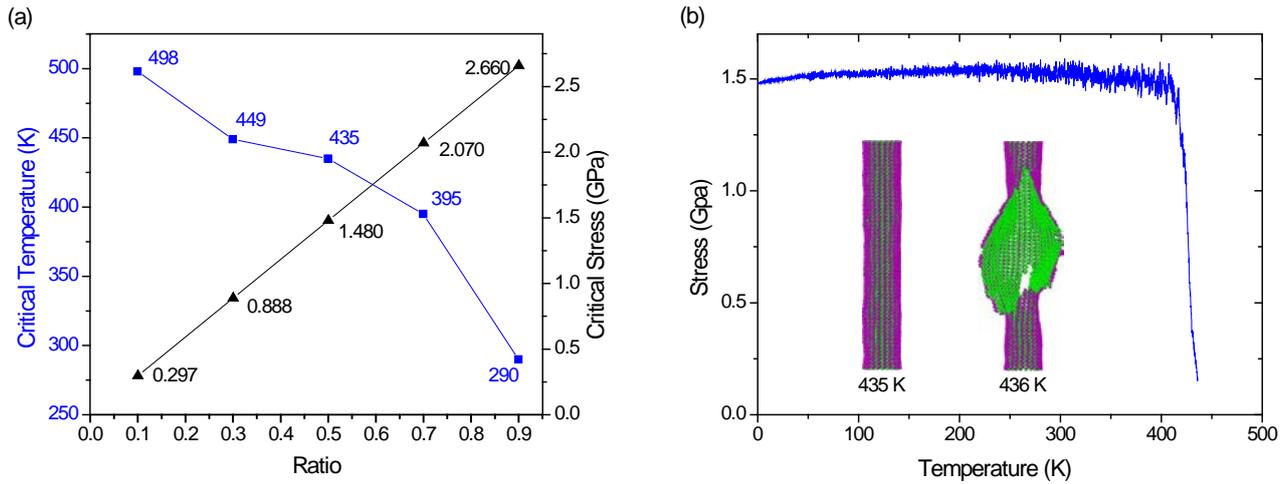

Figure 8. (a) The critical temperature and stress of the BPNT (NA=20, L=20.166 nm) with specified compressed length. Ratio is the specified compressed length over 0.9800 nm. (b) The stress history of the BPNT with 0.5 times of 0.9800nm compressed length during increasing temperature.

From Figure 8a, we find that the BPNT with shorter compressed length can bear higher temperature before failure of the nanostructure. The temperature at which the tube fails is called critical temperature of the tube with specified compressed length. For instance, the critical temperature reaches 498 K when the tube is subjected to 0.1 times of 0.9800 nm of compressed length. When the tube is with the compressed length of 0.9 times of 0.9800 nm, the critical temperature approaches 290K. It means that the present BPNT (with NA=20, L=20.166 nm) has high robustness of environmental temperature. The critical stress of the tube however is still approximately proportional to the compressed length.

In Figure 8b, the stress of the tube is stable during the process of increasing the temperature before reaching 400K. When the temperature is higher than 420K, the stress drops rapidly. This is because the buckling of tube happens. In particular, at 436 K, the tube is broken (see the insert figures and Movie 5). It demonstrates that the critical temperature given in Figure 8a is actually higher than the temperature at which the tube starts buckling. Hence, the critical temperature of 436K, is the upper boundary of the temperature for tube to be fractured. Considering this phenomenon, the nanodevice with such BPNT as a component should work at a temperature lower than the critical temperature.

4. Conclusion

Considering the importance of the mechanical properties of a BPNT on the design of a nanodevice with BPNT as a component, the strength and stability of a BPNT under compression are numerically investigated. From the results, some conclusions are made as follows,

1) The BPNT starts buckling which is followed by fracturing. Before fracturing, the structure has very small axial strain (~5%), hence, the buckling state of a BPNT can be controlled with Euler's buckling formula.

2) For the tube with the same ratio of length over diameter, the critical strain with respect to the strength of a tube decreases with the increasing of the diameter.

3) For the tube with the same cross section, the critical stress (the peak value of stress) is higher for a shorter tube under compression. The critical strain with respect to the strength decreases with the increasing of the length.

4) For the same BPNT under the load with different speeds, the strain of the BPNT, which is still at post-buckling state but not damaged, is higher at lower loading speed.

5) For the same BPNT, it bears higher temperature when under lower compressed length. Hence, the nanodevice with such BPNT as a component should work at a temperature lower than the critical temperature with respect to strength.

Acknowledgement

The financial support from the National Natural-Science-Foundation of China (Grant No.11372100) is acknowledged.

References

Supporting materials:

    Movie 1-NA=14 & [0.73, 0.85]nm compressed length.avi
    Movie 2-NA=20 & [0.93, 1.12]nm compressed length.avi
    Movie 3- a=6 & [0.78, 1.00]nm compressed length.avi
    Movie 4-Relax 0.2ps & [0.31, 1.50]nm compressed length.avi
    Movie 5-Ratio=0.5time & [350, 436]K temperature.avi